\documentclass[a4paper,11pt]{article}
\usepackage{jcappub} 
\usepackage{lineno}
\usepackage[normalem]{ulem}
\usepackage{xcolor}

\title{\boldmath Time-dependent Evolution of Proton Spectra in Supernova Remnants and Their Contribution to Galactic Cosmic Rays}

\author[a,b]{Jun-Yu Shen}
\author[a,b,c]{Hou-Dun Zeng}
\author[a,b]{Qiang Yuan}
\affiliation[a]{Purple Mountain Observatory, Chinese Academy of Sciences, Nanjing 210023, People's Republic of China}

\affiliation[b]{School of Astronomy and Space Science, University of Science and Technology of China, Hefei 230026, People's Republic of China.}

\affiliation[c]{Key Laboratory of Astroparticle Physics of Yunnan Province, Yunnan University, Kunming 650091, People's Republic of China}

\emailAdd{zhd@pmo.ac.cn; yuanq@pmo.ac.cn}

\abstract{
Recent $\gamma$-ray observations indicate that the proton spectra of supernova remnants (SNRs) are well described by broken power laws, with both the spectral break energy, $E_{\mathrm{br}}$, and the low-energy spectral index, $\alpha$, exhibiting systematic evolution with SNR age. The physical origin of these evolutionary trends and their implications for the Galactic cosmic-ray (CR) population remain poorly understood. In this work, we develop the temporal evolution model for protons in SNRs by extending the semi-analytical framework of Zhang \& Fang, in which both the maximum acceleration energy and the injection spectral index evolve with the dynamical evolution of the remnant. The calculated proton spectra reproduce the age-dependent trends of both $E_{\mathrm{br}}$ and $\alpha$ inferred from observations.
We adopt the proton spectrum at the onset of the radiative phase as the source spectrum for Galactic CR propagation and incorporate the intrinsic dispersion of source spectral indices among SNRs. The resulting cumulative Galactic proton spectrum is then calculated within a diffusion model.
The propagated spectrum agrees well with the observed CR proton flux over a broad energy range, particularly above several tens of GeV. Our results provide a self-consistent framework linking the time-dependent evolution of proton acceleration in individual SNRs to the Galactic CR proton spectrum observed at Earth, and further support the long-standing hypothesis that SNRs are the dominant sources of Galactic CR protons below the knee.}

\begin{document}
\maketitle
\flushbottom

\section{Introduction}
\label{sec:intro}
Supernova remnants (SNRs) are widely regarded as the primary accelerators of Galactic cosmic rays (CRs) up to the knee at $\sim3$~PeV \citep{Hillas2005,Blasi2013,Ohira2016}. Over the past decades, increasing observational evidence from radio, X-ray, and $\gamma$-ray observations has provided strong support for this paradigm by revealing efficient particle acceleration at SNR shocks \citep{Helder2012,Yang2014,Eppens2024,Archer2025}. Within the framework of diffusive shock acceleration (DSA), charged particles repeatedly scatter across collisionless shocks and acquire a power-law energy spectrum \citep{Drury1983,Amato2014,Caprioli2024}. Combined with Galactic propagation models, DSA successfully reproduces many of the observed properties of Galactic CRs and remains the leading theoretical framework for their origin \citep{Ptuskin2010,Ptuskin2013,Berezhko2014,Zhang2017}.

Although the SNR origin of Galactic CRs is now well established, the temporal evolution of the accelerated particle spectrum throughout the lifetime of an SNR remains incompletely understood. Since protons constitute nearly 90\% of Galactic CRs, understanding the evolution of the proton spectrum inside SNRs is essential for connecting shock acceleration with the CR spectrum measured at Earth. Using $\gamma$-ray observations of 35 Galactic SNRs, Zeng et al.\cite{Zeng2019} showed that the proton spectrum is well described by a broken power law whose break energy, $E_{\mathrm{br}}$, and low-energy spectral index, $\alpha$, both evolve systematically with SNR age. These observational correlations provide direct empirical constraints on the time dependence of particle acceleration in SNRs and therefore offer an opportunity to test theoretical models of CR production.

Several physical mechanisms have been proposed to explain the observed evolution of the spectral break. As an SNR expands, the shock velocity decreases and the maximum attainable particle energy declines. At the same time, adiabatic expansion, energy-dependent particle escape, and the evolution of magnetic turbulence further modify the high-energy proton spectrum \citep{Ohira2010,Ohira2011,Malkov2011,Celli2019,Zeng2019,Brose2020}. Together, these processes are expected to shift the spectral break toward lower energies as the remnant evolves. However, whether their combined effects can quantitatively reproduce the observed evolution of $E_{\mathrm{br}}$ has not yet been systematically investigated.

The physical origin of the evolution of the spectral index $\alpha$ is even less certain. Within the framework of DSA, the spectral index can be approximately related to an effective compression ratio $r$ characterizing the overall shock modification. Under the test-particle limit, a strong shock yields $r=4$, which corresponds to a spectral index of 2.0. In realistic SNRs, however, $r$ is dynamically modified by multiple physical processes, such as magnetic-field amplification, particle escape, and radiative cooling. For young remnants, magnetic-field amplification driven by streaming instabilities and turbulent dynamo processes enhances the upstream magnetic pressure, thereby reducing $r$ below the canonical value of $4$ and producing softer particle spectra \citep{Bell1978,Bell2004,Xu2016,Xu2017,Vladimirov2006,Caprioli2008,Caprioli2009,Recchia2018,Cristofari2025}. As the remnant ages, the gradual damping of magnetic turbulence, together with the increasing influence of relativistic particles and radiative compression, may increase $r$, resulting in progressively harder spectra \citep{Berezhko1999,Ellison2005,Baring1999,BP2010,Sushch2023}. These theoretical expectations are broadly consistent with the observational trends reported by Zeng et al.\cite{Zeng2019}, but a quantitative comparison between theory and observations is still lacking.

Another important issue concerns the connection between particle acceleration inside individual SNRs and the Galactic CR spectrum. Numerical simulations have shown that a substantial fraction of accelerated particles remain confined within or close to the forward shock during most of the Sedov evolution before gradually escaping into the interstellar medium (ISM) \citep{Reville2009,Fujita2010,Fujita2011,Telezhinsky2012,Celli2019}. Consequently, the proton spectrum released at the end of the SNR evolution carries the integrated imprint of the time-dependent acceleration history. Coupling the released spectrum with Galactic CR propagation therefore provides a natural framework for linking the evolution of individual remnants to the CR proton spectrum observed at Earth.

Motivated by these considerations, we develop a time-dependent model for proton acceleration and evolution in SNRs by extending the semi-analytical framework of Zhang \& Fang \cite{ZF2007}. Guided by the observational correlations reported by Zeng et al.\cite{Zeng2019}, we introduce time-dependent descriptions of both the maximum acceleration energy and the proton injection spectral index. We first calculate the evolution of the proton spectrum throughout the lifetime of an SNR and compare the model predictions quantitatively with the observed age dependence of $E_{\mathrm{br}}$ and $\alpha$. The resulting proton spectrum at the onset of the radiative phase is then adopted as the source term for Galactic CR propagation. By combining the contributions from a population of Galactic SNRs with a spatial distribution and supernova rate, we calculate the CR proton spectrum at Earth and compare it with the latest observations. This work establishes a unified framework that connects the dynamic evolution of SNRs, the temporal evolution of proton acceleration, and the Galactic propagation of CRs within a single self-consistent model.

This paper is organized as follows. Section~2 describes the temporal evolution for protons in SNRs and the Galactic CR propagation framework. Section~3 presents the numerical results and comparisons with observations. The main conclusions and discussion are shown in Section~4.

\section{Model Descriptions}

\subsection{Temporal evolution for protons in an SNR}
To model the temporal evolution of CR protons in SNRs, we build upon the semi-analytical framework developed by \cite{ZF2007}. In the present work, we generalize this framework by introducing time-dependent descriptions of the maximum acceleration energy and the proton injection spectral index, both constrained by recent observational results. The complete model therefore consists of three components: the dynamic evolution of the SNR, proton acceleration at the forward shock, and the subsequent evolution of the proton energy spectrum within the expanding remnant.

 
\subsubsection{SNR evolution}
 In this model, our benchmark SNR is assumed to expand into a homogeneous ambient medium with density $n_{0}$. It will evolve through three phases: the free expansion phase, the Sedov-Taylor
(ST) phase, and the radiation phase. With an explosion energy of $E_{\text{SN}}=10^{51}E_{51}$ erg and an initial shock velocity of $v_{0}$, the ST phase begins at $t_{\text{ST}}\approx2.1\times10^{17}(E_{51}/n_{0})^{1/3}v_{0}^{-5/3}$ yr, while the radiation phase begins at $t_{\text{rad}}\approx4.0\times10^{4}E_{51}^{4/17}n_{0}^{-9/17}$ yr \citep{Blondin1998,Yam2006}, where $n_{0}=\mu n_{\mathrm{ISM}}$. $n_{\mathrm{ISM}}$ is the hydrogen density in the local ISM and $\mu$ is the mean atomic weight of the ISM. During these three phases, the shock velocity $v_{\text{s}}(t)$ is described through \citep{Yam2006}:  
\begin{equation}\label{1}
	v_{\text{s}}(t) = \begin{cases} 
v_0, & t < t_{\text{ST}}, \\
v_0 \left( \frac{t}{t_{\text{ST}}} \right)^{-3/5}, & t_{\text{ST}} \leq t < t_{\text{rad}}, \\
v_0 \left( \frac{t_{\text{rad}}}{t_{\text{ST}}} \right)^{-3/5} \left( \frac{t}{t_{\text{rad}}} \right)^{-2/3}, & t > t_{\text{rad}},
\end{cases}
\end{equation}with $v_0=10^{9}$ cm/s.
Thus, the shock radius at an age of $t$ is $R_{\text{s}}(t)=\int_0^tv_{\text{s}}(t)\mathrm{d}t$. Here, we set $E_{\text{SN}}=1.0\times10^{51}$ erg and $n_{\text{ISM}}=0.5$ cm$^{-3}$, which corresponds to $t_{\text{ST}}\approx 200$ yr and $t_{\text{rad}}\approx 4.8\times10^{4}$ yr.

\subsubsection{Distribution of the shock-accelerated protons} 
Following \cite{ZF2007}, the volume-averaged production rates for the shock-accelerated protons is given by:
\begin{equation}\label{2}
Q_p(E, t) = Q^0_{p} G(t) \left[ E(E + 2m_p c^2) \right]^{-[(\alpha_{p}+1)/2]}(E + m_p c^2) \exp \left( -\frac{E}{E_{\text{max}}(t)} \right),
\end{equation}
where $E$, $E_{\text{max}}(t)$, $m_p$, and $\alpha_{p}$ are the proton kinetic energy, the maximum injection energy, proton mass, and injection spectral index, respectively. The time-dependent function $G(t)$ is defined by:
\begin{equation}\label{3}
G(t) = \begin{cases} 
R_{\text{s}}(t_{\text{ST}})/R_{\text{s}}(t), & t \leq t_{\text{rad}}, \\
0, & t > t_{\text{rad}}.
\end{cases}
\end{equation}

The normalization constant $Q^0_{p}$ is determined through the following equation:
\begin{equation}\label{3.5}
\begin{aligned}
E_{\text{par}}= & \int_{0}^{t_{\text{rad}}} \text{d} t V_{\text{SNR}}(t) \int_{0}^{E_{\text{max}}}\text{d} E E Q_{p}(E, t) ,
\end{aligned}
\end{equation}
where $E_{\text{par}}=10^{50}\,\text{erg}$ is the total kinetic energy contained in protons, and $V_{\text{SNR}}(t)=4\pi R_{\text{s}}^{3}(t)/3$ is the SNR volume.

Unlike \cite{ZF2007}, we adopt a more general description for $E_{\text{max}}(t)$, assuming it increases linearly with SNR age in the free expansion phase and decrease as a power law once the ST phase begins \citep{Gabici2009,Celli2019}, namely
\begin{equation}\label{4}
	E_{\text{max}}(t)=\left\{\begin{array}{ll}
E_{\text{M}}\times\left(t / t_{\text{ST}}\right) & t \leqslant t_{\text{ST}}, \\
E_{\text{M}}\times\left(t / t_{\text{ST}}\right)^{-\delta} & t>t_{\text{ST}},
\end{array}\right.
\end{equation}where $E_{\text{M}}$ is the maximum energy attained at $t = t_{\text{ST}}$, and $\delta$ is a free parameter. Here, $E_{\text{M}}$ and $\delta$ are taken to be 1 PeV and 3.5, respectively \citep{Celli2019,Morlino2021}.

Furthermore, in order to reproduce the observed evolution of the spectral index, $\alpha_p$ is parameterized as linear function of logarithmic SNR age:\begin{equation}\label{5}
\alpha_{p}(t) = 
3.05-0.35\log_{10}(t/\text{yr})
\end{equation}

\subsubsection{Temporal evolution of proton distribution}
During the SNR evolution, the differential proton density $n_p(E,t)$ can be obtained through solving Fokker-Planck equation in energy space \citep{ZF2007}:
\begin{equation}\label{6}
\frac{\partial n_p(E, t)}{\partial t} = -\frac{\partial}{\partial E} \left( \dot{E}_{\text{tot}}(E,t) n_p(E, t) \right) + \frac{1}{2} \frac{\partial^2}{\partial E^2} \left( D_{\text{Coul}}(E, t) n_p(E, t) \right) + Q_p(E, t) - \frac{n_p(E, t)}{\tau_{c}},
\end{equation}where the terms on the right-hand side represent systematic energy losses, diffusion in energy space, particle injection function, and catastrophic energy loss. 

In Equation (\ref{6}), $\dot{E}_{\text{tot}}$ is the total energy loss rate for protons, including the Coulomb energy loss rate $\dot{E}_{\text{Coul}}$ and the adiabatic loss rate $\dot{E}_{\text{ad}}$. The expression for $\dot{E}_{\text{Coul}}$ is \citep{Sturner1997}:
\begin{equation}\label{7}
   \dot{E}_{\text{Coul}}(E,t) = -\frac{3}{2} \sigma_{\text{T}} m_e c^3 n_{\text{SNR}}\frac{1}{\beta_p}\lambda(t) \left( \frac{m_e }{m_p} \right) \left[\left( \frac{m_p }{m_e} \right)\psi(t)-\psi^{\prime}(t)\right],
\end{equation}where
\begin{equation}\label{8}
   \psi(t)=\frac{2}{\sqrt{\pi}}\int^{x(t)}_{0}\mathrm{d}y\,\sqrt{y}\,\mathrm{exp}(-y),
\end{equation}$\psi^{\prime}(t)=\mathrm{d}\psi/\mathrm{d}x$, $x(t)=m_ev_p^2/2kT_e(t)$, and the Coulomb
logarithm $\lambda(t)\sim30$. $\sigma_{\text{T}}$, $m_{e}$, and $c$ are the Thomson cross section, electron mass, and speed of light, respectively. $\beta_p$ is the proton velocity $v_p$ divided by $c$. $n_{\text{SNR}}$ symbolizes the proton density inside the SNR and $T_e(t)$ the electron temperature. The adiabatic loss rate for protons is expressed by:
\begin{equation}\label{9}
\dot{E}_{\text{ad}}(E) = -\frac{E}{R_{\text{s}}} \frac{\text{d}R_{\text{s}}}{\text{d}t}.
\end{equation}

The Coulomb diffusion coefficient $ D_{\text{Coul}}(E, t)$ is:
\begin{equation}\label{10}
    D_{\text{Coul}}(E, t) = 3\sigma_{\text{T}} m_e c^3\frac{n_{\text{SNR}}}{\beta_p} \lambda(t) kT_e(t) \psi(t).
\end{equation}

In the context of catastrophic losses, the relevant processes comprise pion-producing collision and particle escape. The corresponding timescales are written as:
\begin{equation}\label{11}
    \tau_{\text{pion}}=1/(\sigma_{pp} n_{\text{SNR}}\beta_{p}c),
\end{equation}where $\sigma_{pp}$ is the inelastic cross section for p-p interaction \citep{Kelner2006}, and
\begin{equation}\label{12}
    \tau_{\text{esc}}=\frac{R_{\text{s}}^{2}}{D(E)},
\end{equation}with $D(E)$ standing for the spatial diffusion coefficient in SNRs \citep{Fatuzzo2006}.

\subsection{Propagation of protons released by SNRs}
The time-dependent acceleration model described above yields the proton energy distribution confined within an SNR throughout its evolution. Once the remnant enters the radiative phase, efficient particle acceleration is assumed to cease, and the accumulated proton spectrum is released into the ISM. In this work, we adopt the proton spectrum at the onset of the radiative phase ($t=t_{\rm rad}$) as the source injection spectrum for Galactic CR propagation. By coupling this source spectrum with a diffusion model for CR transport in the Galaxy, we calculate the cumulative proton flux contributed by the Galactic SNR population and compare the predicted spectrum with observations at Earth.

\subsubsection{Propagation model}
To characterize the proton propagation in the Galaxy, we utilize a simple diffusion model which has been broadly applied in the study of CR transport \citep{Kobayashi2004,Delahaye2009,Ohira2011}. Within this framework, CRs are confined in a cylindrical diffusive halo with a radius of $R = 20$ kpc and a height of 2$L$. Their density vanishes at the boundaries, i.e.,  $N(|z| = L,r) = N(z,r = R) = 0$, where $r$ and $z$ are the radial and vertical distances from the Galactic center, respectively. The CR proton propagation in this halo is governed by a diffusion equation\footnote{For simplicity, we neglect the energy losses, reacceleration, inelastic collision, and convection for protons during propagation.} \citep{Delahaye2009}: 

\begin{equation}\label{13}
\frac{\partial N(E)}{\partial \tau}-K(E) \Delta N=q_{p}(E, \tau, r, z),
\end{equation}
 where $K(E)$, and $q_{p}(\tau, r, z)$ are the Galactic diffusion coefficient and the source function, respectively. Under this model, CR diffusion is homogeneous and isotropic with $K(E)=\beta_p K_0(E/1\ \mathrm{GeV})^{\delta_{D}}$, where $K_0$ and $\delta_{D}$ are constant parameters. Here, we adopt $K_0=0.0112$ kpc$^{2}$/Myr, $\delta_{D}=0.62$, and $L=4$ kpc \citep{Delahaye2009}. Regarding the source function, the final proton spectrum in our benchmark SNR is employed as $q_{p}(E, \tau, r, z)$, which represents the spectrum of protons liberated by an SNR located at $(r,z)$ from the Galactic center at $\tau$ years ago.

 \subsubsection{The Green function}
 The Green function for Equation (\ref{13}) is given by:
 \begin{equation}\label{15}
G(\tau, r, z)=\frac{\theta(\tau)}{4 \pi K_{0} \tau} \exp \left(-\frac{r^{2}}{\lambda_{D}^{2}}\right) \times G^{1 D}(z, \tau),
\end{equation}where $\theta(\tau)$ is the Heaviside function and \(\lambda_D\equiv\sqrt{4K_0\tau}\) is the diffusion length. In order to describe the $z-$directional propagation, the corresponding Green function $G^{1 D}(z, \tau)$ is considered for two regimes \citep{Lavalle2007}:

(i) if $\lambda_{D}$ is less than $L$, the electrical image formula is utilized \citep{BE1999}:
\begin{equation}
G^{1D}(z,\tau) = \sum_{n=-\infty}^{+\infty} (-1)^n \, \frac{\theta(\tau)}{\sqrt{4\pi K_0\,\tau}} \exp\left(-\frac{z_{n}^2}{4 K_0\,\tau}\right),
\label{16}
\end{equation}
where $z_n = 2L\,n$.

(ii) if  $\lambda_{D}$ is greater than $L$, the expression is analogous to the solution of Schrödinger equation in an infinitely deep square potential: 
\begin{equation}
G^{1D}(z,\tau) = \frac{1}{L}\sum_{n=1}^{+\infty} \left[ e^{-K_0 k_n^2\,\tau}\, \phi_n(0)\,\phi_n(z) + e^{-K_0k_{n}^{\prime2}\,\tau}\, \phi'_n(0)\,\phi'_n(z) \right],
\label{17}
\end{equation}
where
\[
\phi_n(z) = \sin\left[k_n\Bigl(L - |z|\Bigr)\right], \quad k_n = \frac{\left(n-\frac{1}{2}\right)\pi}{L} \ (\mathrm{even}),
\]
\[
\phi'_n(z) = \sin\left[k'_n\Bigl(L - z\Bigr)\right], \quad k'_n = \frac{n\pi}{L} \ (\mathrm{odd}).
\]

 \subsubsection{The total proton flux}
The spatial distribution of SNRs in our Galaxy is \citep{Yuan2012}:
\begin{equation}\label{18}
f(r,z) \propto \left(\frac{r}{r_\odot}\right)^{\alpha_r} \exp\left(-{\beta_r} \frac{r - r_\odot}{r_\odot} - \frac{|z|}{z_0}\right),
\end{equation}
where $r_{\odot}=8.5$ kpc is the solar Galactocentric distance and $z_0=0.2$ kpc is the scale height for SNR distribution. In this work, $\alpha_r$ and $\beta_r$ are set as 1.09 and 3.87, respectively \citep{Green2015,Ranasinghe2022}. The positions of the SNRs in our Galaxy will be randomly taken according to Equation (\ref{18}).

If we choose our Earth as the origin of the coordinate system, SNR positions should be transformed as:
\begin{equation}\label{19}
\begin{cases}
r^{\prime} = \sqrt{r^{2}+r^{2}_{\odot}-r_{\odot}\,r\,\text{cos}\theta}\\[1ex]
z^{\prime} = -z,
\end{cases}
\end{equation}where $\theta$ is a random angle ranging from 0$^{\circ}$ to 180$^{\circ}$. 

Consequently, the total proton flux observed at the Earth is given by \citep{Delahaye2009}: 
\begin{equation}
\phi_{p}(E) = \frac{c\,\beta_p}{4\pi}\, \sum^{N_{\text{SNR}}}_{i=1} \; G(\tau_{i}, r^{\prime}_{i}, z^{\prime}_{i}) \, q_{p}(E, \tau_{i}, r^{\prime}_{i}, z^{\prime}_{i}),
\label{20}
\end{equation}where $N_{\text{SNR}}$ is the total number of SNR under a Galactic supernova rate of one per century \citep{van1991}. Concerning solar modulation at low energies ($E<30$ GeV), we employ the force-field approximation with a modulation potential $\Phi=620$ MV \citep{Gleeson1968,Aguilar2015}.

\section{Results}

\begin{figure*}[htbp]
	\centering
	\includegraphics[width = 0.51\linewidth]{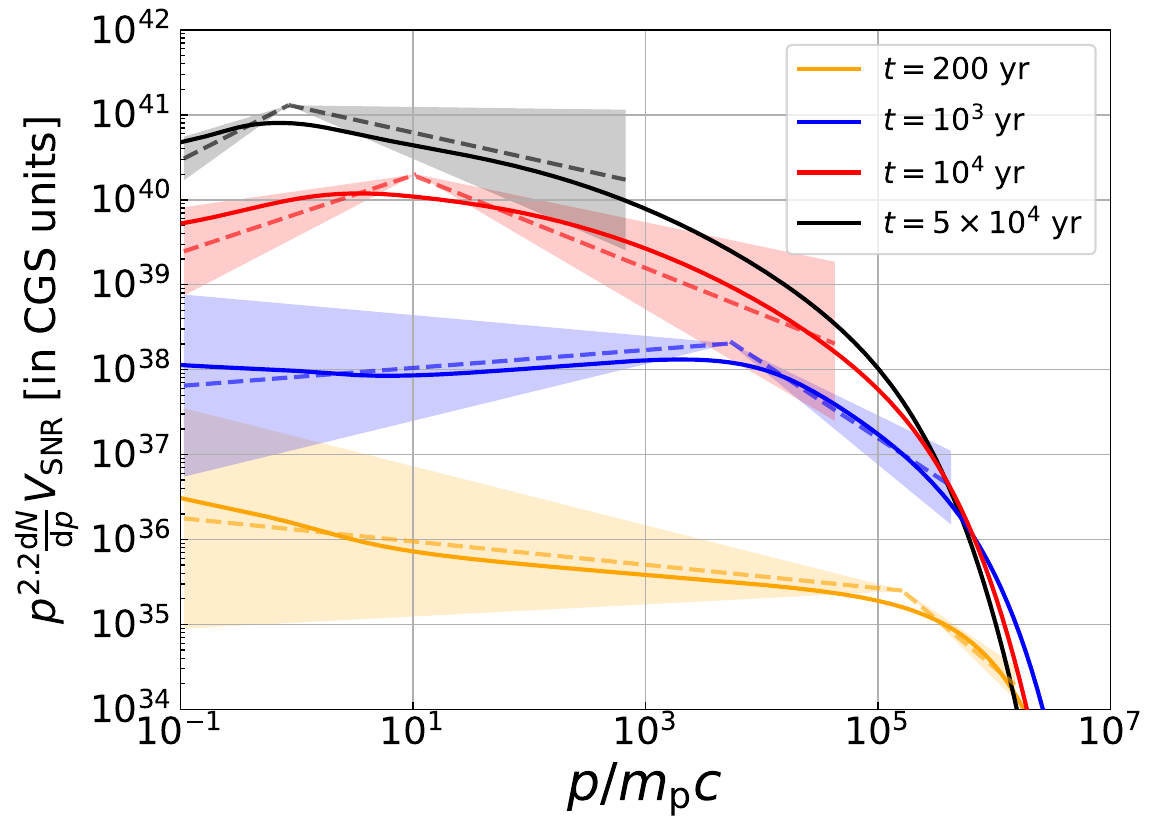}
    \includegraphics[width = 0.48\linewidth]{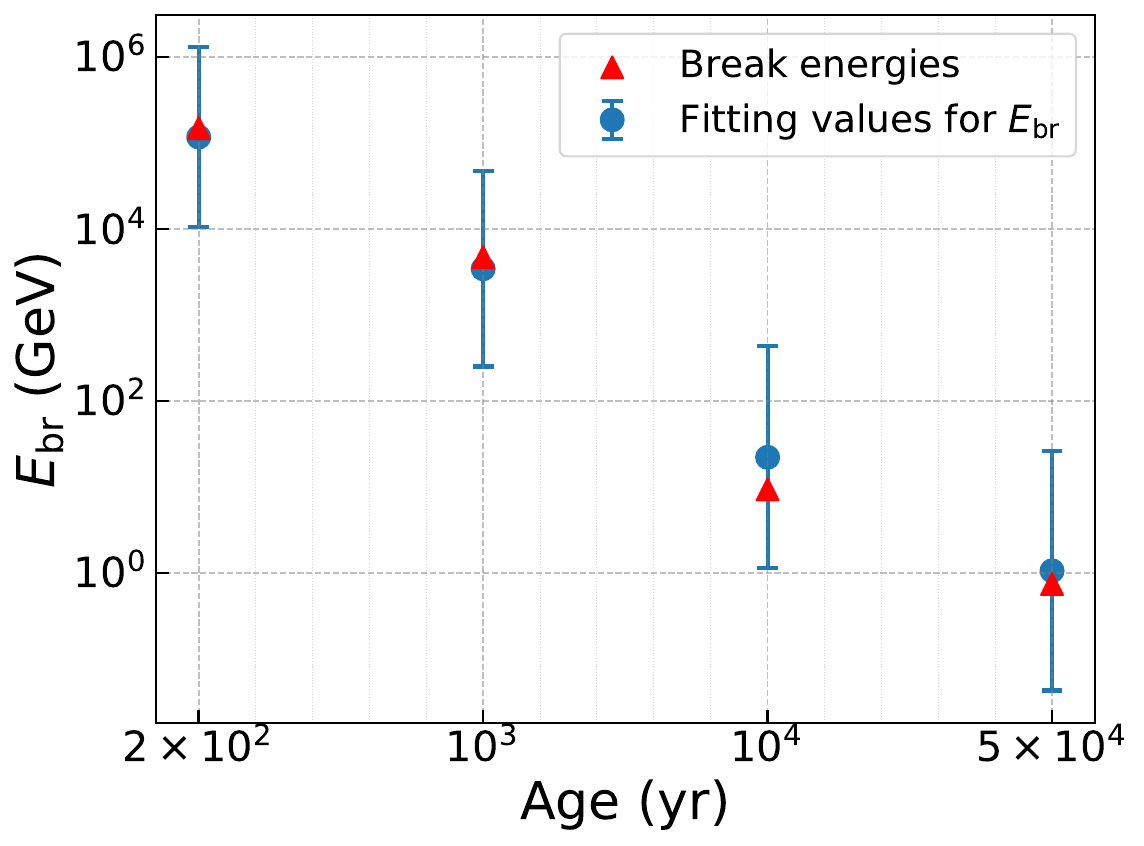}
	\caption{Comparison of model-predicted proton spectra with observational constraints. Left panel: solid curves show spectra at 200, $10^{3}$, $10^{4}$, and $5\times10^{4}$ yr. Dashed lines and shaded bands represent broken power-law fits with slopes 
    $\alpha$ (low energies) and $\alpha+1.0$ (high energies) from Zeng et al.~\cite{Zeng2019}, with $1\sigma$ uncertainties in $\alpha$.
    Right panel: break energies $E_{\mathrm{br}}$ from the calculated spectra in the left panel (triangles) versus observational inferences (vertical error bars, $1\sigma$). Detailed observational constraints are presented in Appendix~\ref{app1}.}
	\label{fig:2}
\end{figure*}

Based on the time-dependent acceleration model described in Section~2.1, we calculated the evolution of the proton energy distribution inside an SNR throughout its lifetime. The left panel of Figure~\ref{fig:2} displays the proton spectra at representative evolutionary stages of 200, $10^{3}$, $10^{4}$, and $5\times10^{4}$ yr, shown as the yellow, blue, red, and black solid curves, respectively. The corresponding dashed curves represent broken power-law spectra derived from the empirical relations of Zeng et al.\cite{Zeng2019}, with shaded bands indicating the $1\sigma$ uncertainties in the spectral index. The right panel of Figure~\ref{fig:2} compares the break energies $E_{\mathrm{br}}$ extracted from the calculated spectra in the left panel with the observational constraints, where the vertical error bars denote the inferred values and their $1\sigma$ uncertainties (see Appendix~\ref{app1}).

The calculated spectra exhibit a pronounced broken power-law shape whose spectral index and break both evolve systematically with SNR age. Overall, the predicted spectra are in excellent agreement with the observationally inferred spectral indices and break energies, demonstrating that the time-dependent acceleration model successfully reproduces the observed evolution of proton spectra in Galactic SNRs.

A prominent feature of Figure~\ref{fig:2} is the continuous decrease of the break energy, from approximately $100~\mathrm{TeV}$ at $t\simeq200$ yr to about $1~\mathrm{GeV}$ at $t=5\times10^{4}$ yr. This behavior is closely associated with the temporal evolution of the maximum acceleration energy, $E_{\max}(t)$. As described by Equation~(\ref{4}), $E_{\max}$ reaches its maximum value of approximately $1~\mathrm{PeV}$ near the onset of the ST phase and subsequently decreases rapidly as the shock decelerates. Consequently, the injection of newly accelerated particles at the highest energies becomes progressively suppressed during the subsequent evolution of the remnant. Meanwhile, adiabatic expansion continuously reduces the energies of particles confined within the remnant, leading to a gradual softening of the high-energy spectrum and a systematic shift of the spectral break toward lower energies.

Upon entering the radiative phase ($t=5\times10^{4}$ yr), the efficiency of particle acceleration at the shock becomes negligible, and the SNR no longer contributes significantly to the production of newly accelerated CRs. The particles accumulated during the previous evolutionary stages are subsequently released into the ISM and propagate diffusively through the interstellar space within Galactic as individual CR sources. We therefore adopt the proton spectrum obtained at the onset of the radiative phase as the injection spectrum, $q_{p}(E,\tau,r,z)$, in the Galactic diffusion equation (Equation~\ref{13}).
Observations indicate substantial diversity in the spectral properties of individual SNRs, which may arise from variations in ambient density, explosion energy, magnetic-field amplification, and evolutionary stage. To account for this intrinsic population dispersion, we introduce a Gaussian-distributed modification to the spectral index of each SNR. Specifically, the injection spectrum is expressed as
\begin{equation}
q_{p}^{\prime}(E,\tau,r,z)=q_{p}(E,\tau,r,z)E^{-\sigma_{p}},
\end{equation}
where $\sigma_{p}$ follows a Gaussian distribution with a mean value of zero and a standard deviation of $\sigma_{p}=0.31$.\footnote{The adopted value of $\sigma_{p}$ corresponds to the width of the confidence interval of the spectral index $\alpha$ at $t=5\times10^{4}$ yr (see Figure~\ref{fig:5}).}
This treatment allows us to incorporate the observed diversity of SNR proton spectra into the Galactic CR population synthesis. 

\begin{figure*}[htbp]
	\centering
	\includegraphics[width = 0.65\linewidth]{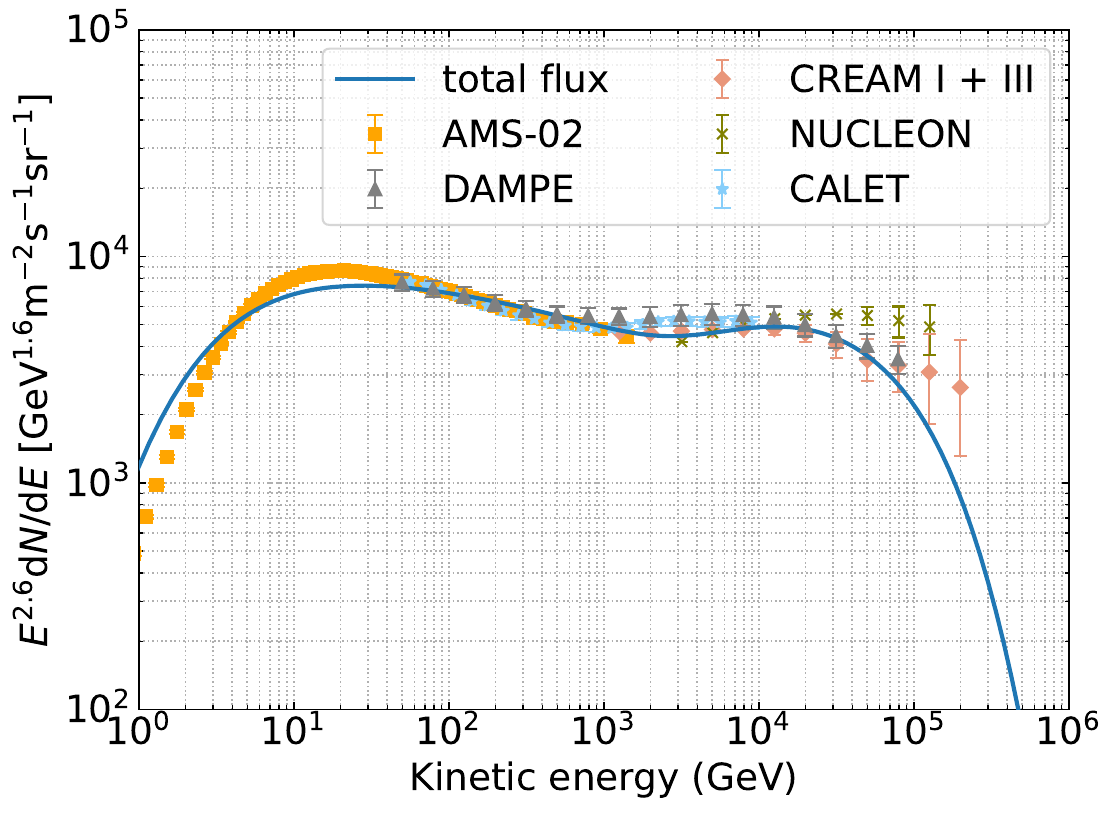}
    \caption{Calculated proton spectrum after solar modulation. References of the data are: AMS-02 \citep{Aguilar2015}, CREAM I + II \citep{Yoon2017}, CALET \citep{Adriani2019}, DAMPE \citep{An2019}, and NUCLEON \citep{Grebenyuk2019}.}	
	\label{fig:4}
\end{figure*}

Using the propagation framework described in Section~2.2, we calculate the cumulative proton flux at Earth. Figure~\ref{fig:4} presents the resulting Galactic proton spectrum after propagation, together with measurements from various CR experiments for comparison. Two characteristic features of the observed Galactic proton spectrum are reproduced by the model.  
First, the mild hardening near $\sim200$--300 GeV arises naturally from the dispersion of the source injection spectra \citep{Yuan2011}, without introducing an additional source population or a break in the diffusion coefficient. 
Second, the rapid softening above $\sim10^{5}$ GeV reflects the finite maximum energy attainable in SNR shocks during their evolution. 
At energies below several tens of GeV, the predicted flux falls below the observations. This discrepancy is expected because the present propagation model neglects several processes that become important at low energies, including diffusive reacceleration, Galactic convection, ionization and Coulomb losses, and the detailed effects of solar modulation. Incorporating these physical processes is expected to improve the agreement in the low-energy regime but is beyond the scope of this work. 
Overall, the good agreement between the calculated and observed spectra at intermediate and high energies indicates that the proton spectra derived from SNRs provide a physically plausible description of the dominant Galactic CR proton population below the knee. This conclusion is consistent with previous theoretical studies based on DSA and Galactic CR propagation \citep{Ptuskin2010,Ptuskin2013,Berezhko2014,Zhang2017}. Taken together, our results establish a self-consistent framework that connects the temporal evolution of proton acceleration inside SNRs with the large-scale Galactic CR spectrum observed at Earth.
 
\section{Conclusions and Discussion}

In this work, we develop a time-dependent model for the evolution of CR protons accelerated in SNRs by extending the semi-analytical framework of Zhang \& Fang \cite{ZF2007}. Different from the original model, both the maximum acceleration energy and the injection spectral index are treated as time-dependent quantities, allowing the particle acceleration process to evolve self-consistently with the dynamical evolution of the SNR. The resulting proton spectra reproduce the observed evolutionary trend reported by Zeng et al.\cite{Zeng2019}, including the progressive decrease of the break energy and the spectral hardening with increasing SNR age.

Using the evolved proton spectra as the source term for Galactic CR propagation, we further calculated the cumulative proton spectrum at Earth within a diffusion model that incorporates the Galactic spatial distribution of SNRs and a supernova rate of one event per century. By introducing a Gaussian dispersion in the source spectral indices to account for the intrinsic diversity of individual remnants, the propagated proton spectrum successfully reproduces the observed Galactic proton flux over a broad energy range, particularly above several tens of GeV. 

These results provide additional support for the long-standing hypothesis that SNRs are the dominant sources of Galactic CRs below the knee. More importantly, the present work establishes a self-consistent framework that connects the time-dependent acceleration of particles inside individual SNRs with the large-scale Galactic CR spectrum observed at Earth.

Although the present model successfully reproduces both the temporal evolution of proton spectra in SNRs and the Galactic proton spectrum observed at Earth, several simplifying assumptions have been adopted. In particular, we assume a homogeneous ambient medium and employ a simplified description of Galactic CR propagation. More importantly, the temporal evolution of the proton spectral parameters is introduced phenomenologically rather than being derived self-consistently from nonlinear DSA. Nevertheless, the inferred evolutionary trends are broadly consistent with recent theoretical and numerical studies, providing valuable insight into the underlying physical processes.

One of the principal results of this work is the systematic decrease of the proton break energy, $E_{\mathrm{br}}$, with increasing SNR age. In our model, this evolution phenomenologically is the combined effects of continuous adiabatic cooling and the gradual softening of the high-energy proton spectrum. From a physical perspective, however, the decline of $E_{\mathrm{br}}$ is likely associated with the evolution of magnetic turbulence around the shock. Recent studies have shown that the damping or evanescence of Alfvén waves during the late evolutionary stages of SNRs reduces the level of magnetic turbulence upstream of the shock, thereby increasing the diffusion coefficient and prolonging the particle acceleration timescale \citep{Brose2020,Das2024}. As a consequence, the maximum energy attainable by shock acceleration decreases, while the highest-energy particles escape increasingly efficiently from the acceleration region. These effects naturally lead to a progressive reduction of the spectral break energy as the remnant evolves.

Our model also predicts a gradual hardening of the low-energy proton spectrum with increasing SNR age, which is described by a decreasing injection spectral index, $\alpha_p(t)$. Within the framework of DSA, the spectral index is sensitive to the effective shock compression ratio, $r$, implying that a harder spectrum corresponds to a larger $r$. A number of nonlinear DSA simulations have indeed suggested that $r$ evolves with the aging of an SNR \citep{Ksenofontov2004,Ellison2005,Berezhko2006,Lee2012}. While the canonical value $r=4$ ($\alpha_p=2$) applies to strong shocks in the test-particle limit \citep{Blasi2013}, observations indicate that several young SNRs exhibit softer spectra ($\alpha>2$), corresponding to $r<4$, whereas many evolved remnants display harder spectra with $\alpha<2$, implying $r>4$. Such observational trends are qualitatively consistent with the temporal evolution adopted in the present work. 

\begin{figure*}
	\centering
	\includegraphics[width = 0.65\linewidth]{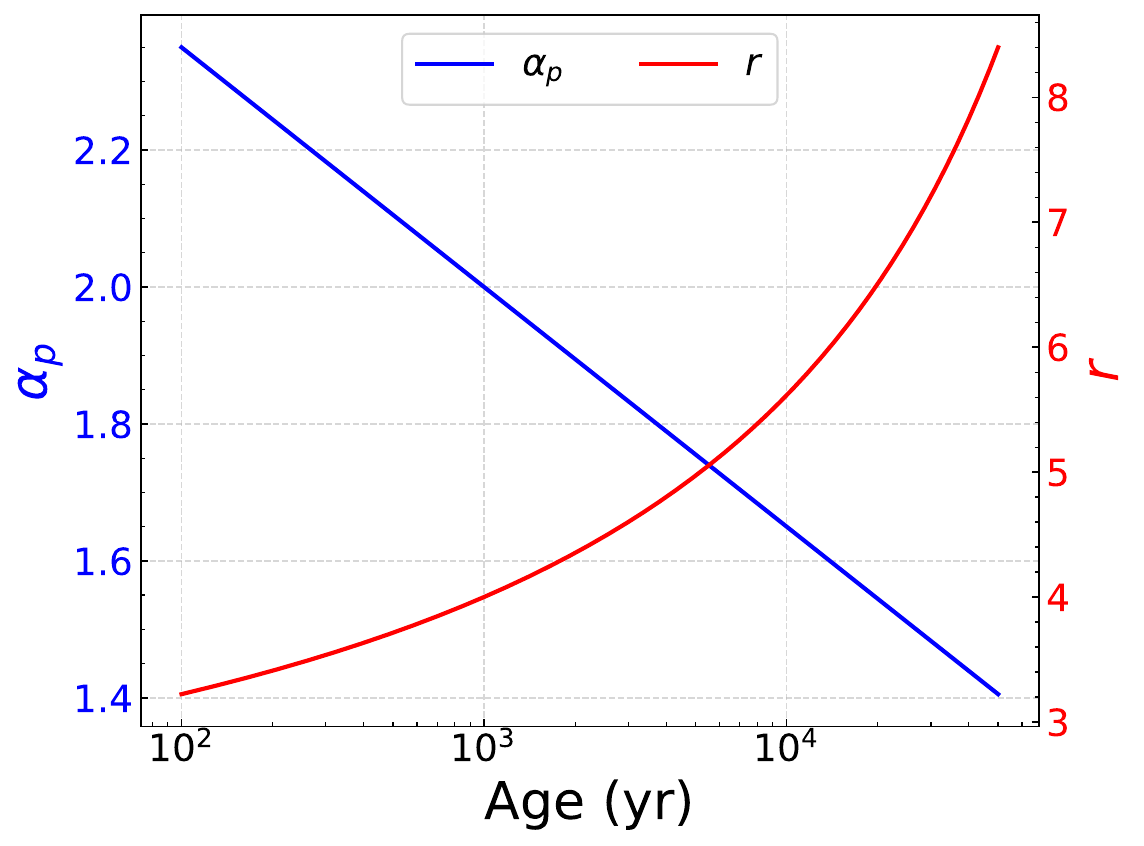}	
	\caption{The injection spectral index $\alpha_{p}$ and effective shock compression ratio $r$ vs. SNR age.}
	\label{fig:1}
\end{figure*}

A plausible physical explanation for the evolution of $r$ is related to the evolution of magnetic turbulence. During the early stages of SNR evolution, efficient magnetic-field amplification upstream of the forward shock can be driven by turbulent dynamo processes \citep{Xu2016,Xu2017}, resonant streaming instability \citep{Bell1978}, and non-resonant Bell instability \citep{Bell2004}. The amplified magnetic pressure reduces the compressibility of the upstream plasma, thereby producing $r$ below the canonical value \citep{Vladimirov2006,Diesing2021}. As the SNR evolves, however, Alfvén-wave damping weakens the magnetic amplification \citep{Brose2016,Brose2021}, gradually diminishing its influence on the shock structure. At the same time, the production of relativistic particles and the escape of superthermal particles increase the downstream compressibility \citep{Blondin2001}, allowing $r$ to evolve from $<4$ towards $>4$. This evolution provides a natural explanation for the gradual hardening of the low-energy proton spectrum inferred from observations. Quantitatively, the evolution of \(r\) is linked to that of \(\alpha_p\) through $r(t)=(2+\alpha_{p}(t))/(\alpha_{p}(t)-1)$ \citep{Recchia2018}. With regard to Equation~\eqref{5}, the resulting $\alpha_{p}(t)$ and $r(t)$ are plotted versus SNR age in Figure \ref{fig:1}.

The proton spectrum measured near the Earth exhibits a spectral hardening at $\sim$200--300 GeV followed by a gradual softening above $\sim$15 TeV, forming a characteristic bump-like structure \citep{Yuan2026}. Our result suggests that the time-dependent evolution of SNR proton spectra alone may provide a viable explanation without invoking additional source populations or modifications to the Galactic propagation model.
Nevertheless, several alternative interpretations of this spectral structure have been proposed. From the perspective of source physics, the bump may originate from nonlinear effects in DSA, which introduce curvature into the accelerated particle spectra \citep{Ptuskin2013}, or from the superposition of proton populations released at different evolutionary stages of SNRs \citep{Zhang2017}. On the other hand, propagation-related processes may also contribute, including spatially dependent diffusion in the Galaxy \citep{Tomassetti2012} and rigidity-dependent transport associated with the excitation and damping of magnetohydrodynamic turbulence \citep{Chernyshov2022,DAMPE2026}. In addition, local contributions from nearby CR sources or reacceleration by local shocks have been suggested as possible mechanisms capable of producing similar spectral features \citep{Savchenko2015,Liu2019,Yuan2026,Malkov2021,Malkov2022,Hu2026}.

\appendix
\renewcommand{\thefigure}{\thesection\arabic{figure}}
\setcounter{figure}{0}

\renewcommand{\thetable}{\thesection\arabic{table}}
\setcounter{table}{0}

\section{Empirical Fits to the Spectral Evolution of Supernova Remnants}
\label{app1}

To quantify the observed evolution of the proton spectral properties with SNR age, we reanalyze the sample compiled by Zeng et al.\cite{Zeng2019}. The break energy, $E_{\mathrm{br}}$, and the low-energy spectral index, $\alpha$, are fitted independently as linear functions of the logarithmic SNR age. The resulting best-fitting relations are

\begin{equation}
\label{eq:appendix_fit}
\begin{aligned}
\log_{10}\left(\frac{E_{\mathrm{br}}}{\mathrm{GeV}}\right)
&=
(-2.19 \pm 0.23)
\log_{10}\left(\frac{\mathrm{Age}}{\mathrm{yr}}\right)
+
(10.11 \pm 0.90),
\\
\alpha
&=
(-0.35 \pm 0.05)
\log_{10}\left(\frac{\mathrm{Age}}{\mathrm{yr}}\right)
+
(3.14 \pm 0.18),
\end{aligned}
\end{equation}
where the quoted uncertainties correspond to the $1\sigma$ statistical errors of the linear regression.

These empirical relations provide quantitative constraints on the temporal evolution of the proton spectrum inferred from $\gamma$-ray observations of Galactic SNRs. Throughout this work, they are adopted as observational benchmarks for evaluating the performance of our time-dependent acceleration model. In particular, the fitted relation for $\alpha$ is also used to estimate the intrinsic dispersion of the source spectral index, which is incorporated into the Galactic population synthesis discussed in Section~3.

\begin{figure*}[htbp]
    \centering
    \includegraphics[width=0.49\linewidth]{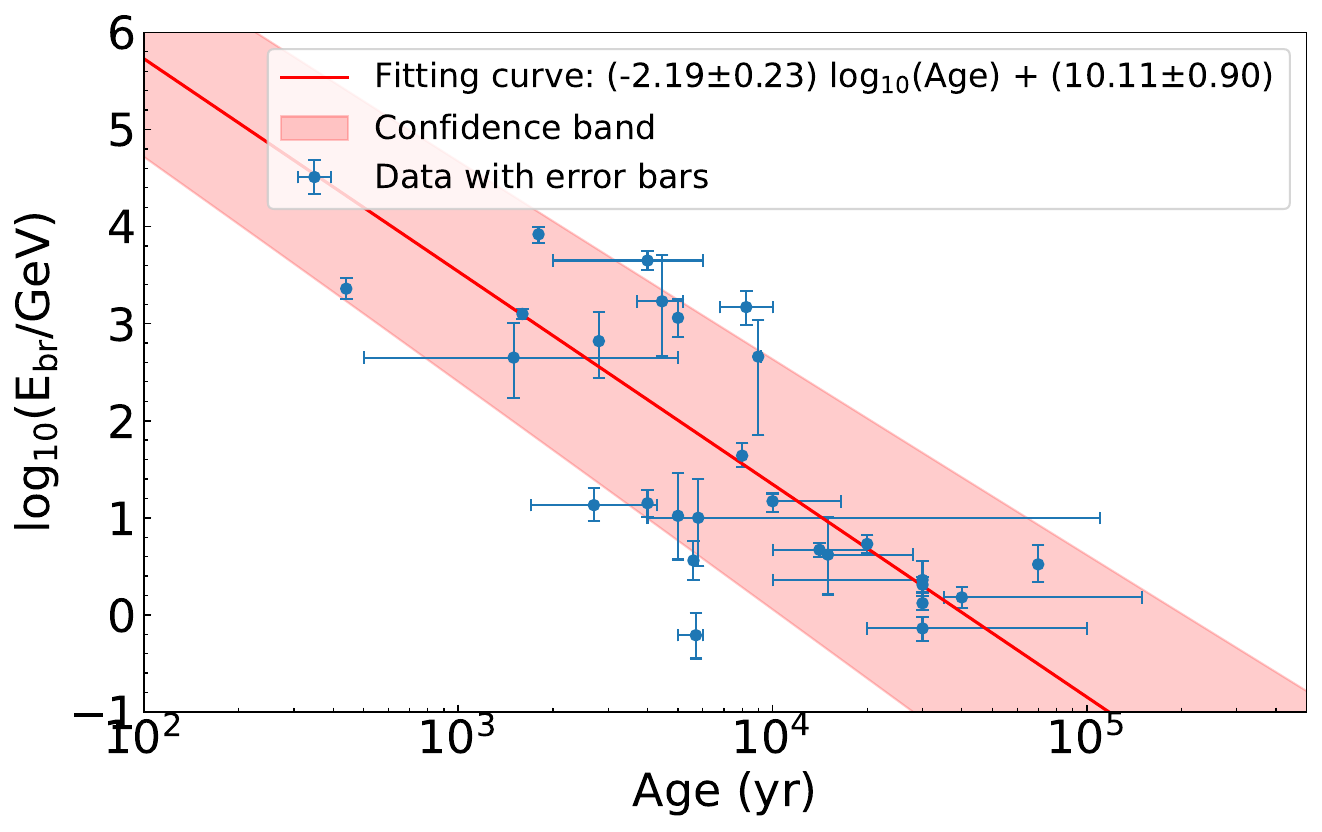}
    \includegraphics[width=0.49\linewidth]{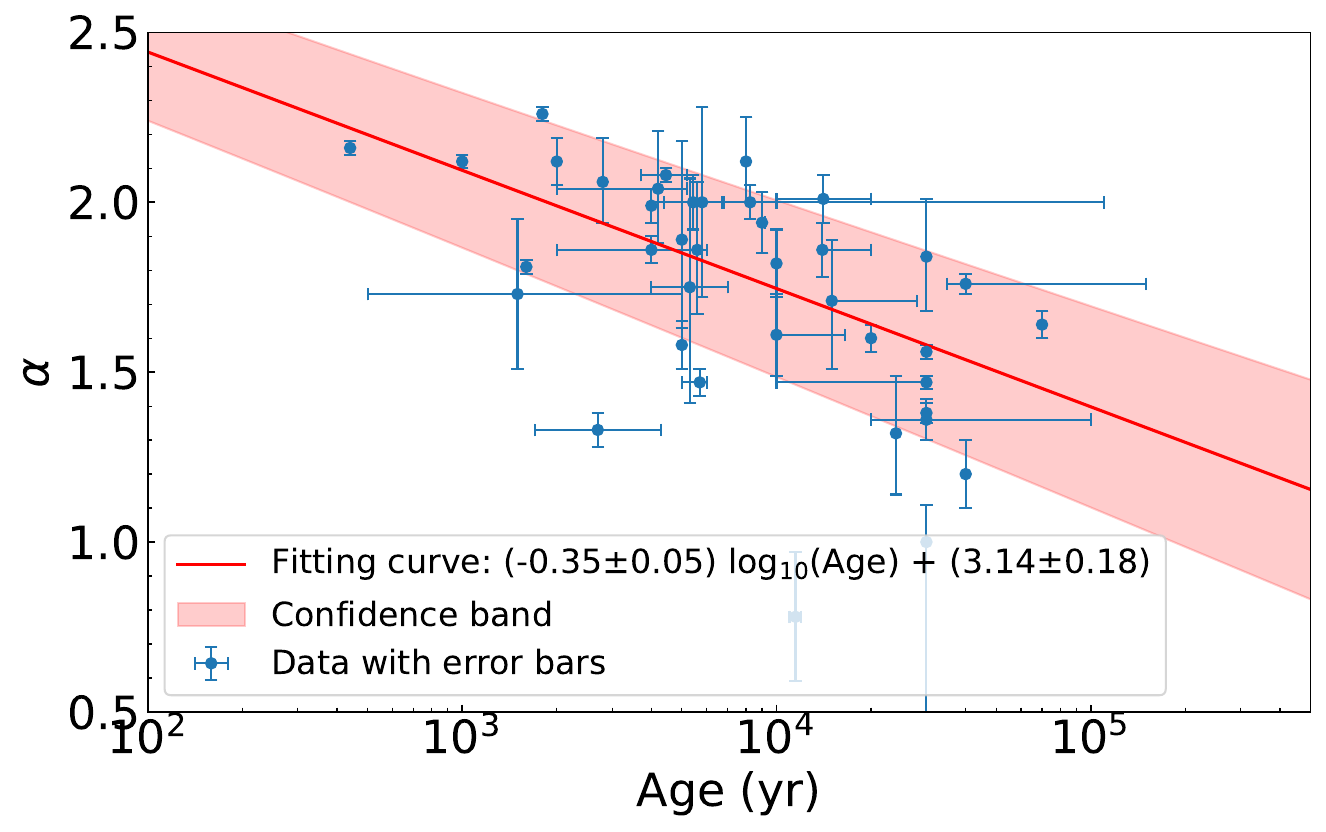}
    \caption{
    Empirical correlations between the proton spectral parameters and the age of Galactic SNRs, obtained by fitting the observational sample compiled by Zeng et al.\cite{Zeng2019}. The left panel shows the evolution of the spectral break energy, $E_{\mathrm{br}}$, while the right panel presents the evolution of the low-energy spectral index, $\alpha$. The solid lines denote the best-fitting linear relations in logarithmic space given by Equation~(\ref{eq:appendix_fit}), and the shaded regions represent the corresponding $1\sigma$ confidence intervals. These empirical trends serve as observational benchmarks for the time-dependent proton acceleration model developed in this work.
    }
    \label{fig:5}
\end{figure*}

Figure~\ref{fig:5} presents the best-fitting relations together with their corresponding confidence bands.

\acknowledgments

This work is supported by the National Natural Science Foundation 
of China (Nos. 12220101003, 12573053, 12273114, and 12322302),
the Project for Young Scientists in Basic Research of the Chinese 
Academy of Sciences (No. YSBR-061), and the Natural Science Foundation 
for General Program of Jiangsu Province of China (No. BK20242114).




\bibliographystyle{JHEP}
\bibliography{biblio}{}

\end{document}